# From Cancer Drivers to Cancer Keepers: Paradigm Shift and Clinical Implications

Xizhe Zhang[1,2*] and Weixiong Zhang[3*]

1. Early Intervention Unit, Department of Psychiatry, The Affiliated Brain Hospital of Nanjing Medical University, Nanjing, Jiangsu 210029, China
2. School of Biomedical Engineering and Informatics, Nanjing Medical University, Nanjing, Jiangsu 210001, China
3. Department of Health Technology and Informatics, Department of Data Science and Artificial Intelligence, Department of Computing, The Hong Kong Polytechnic University, Hong Kong 999077, China

*: Correspondence: XZ: zhangxizhe@njmu.edu.cn, WZ: weixiong.zhang@polyu.edu.hk,

**Abstract:**

Cancer research has traditionally focused on identifying driver genes, those with mutations that initiate tumorigenesis. The *Cancer Driver Gene* (CDG) paradigm, further supported by the observation of oncogene addiction in tumors, has successfully guided the development of targeted therapies. However, the limitations of this driver-centric view – highlighted by the broad emergence of frequent therapeutic resistance, the presence of driver mutations in healthy tissues or individuals, and the lack of identifiable drivers in many tumors – call for a shift in perspective and clinical practice. The latest network controllability perspective on cancer cells introduced the concept of *Cancer Keeper Genes* (CKGs) and a CKG-based paradigm for cancer therapeutics. The new concept encompasses the concept of non-oncogene addiction, emphasizing reliance on non-mutated pathways crucial for maintaining oncogenic cellular states. Here, we explore the transition towards a system-level understanding of cancer based on the CKG paradigm, emphasizing the essential role of genes required for tumor *maintenance*, irrespective of their initiating function or mutational capacity. We discuss clinical implications of this paradigm shift, highlighting the progress made so far and potential of targeting non-driver CKGs – genes involved in processes like DNA damage response, proteostasis, and metabolism – as a promising strategy to overcome therapeutic challenges and achieve more durable cancer control. Targeting these maintenance vulnerabilities represents a critical evolution in precision oncology, moving towards therapies designed to dismantle the networks sustaining malignancies.

## 1. Introduction

For decades, the dominant paradigm in cancer research has centered on the identification of somatic mutations in specific genes, referred to as *Cancer Driver Genes* (CDGs), that confer a selective growth advantage and initiate the process of tumorigenesis[1,2]. This perspective, often encapsulated in the phrase "cancer as a genetic disease"[1,2], posits that the accumulation of mutations in oncogenes and tumor suppressors is the primary force underlying neoplastic transformation[3,4]. This framework has been immensely productive, leading to the cataloging of hundreds of putative driver genes across various cancer types[5] and providing the rationale for developing molecularly targeted therapies aimed at the products of these mutated genes[6]. The success of therapies targeting specific driver alterations, underpinned by the concept of oncogene addiction[7], seemed to validate this driver-centric view.

However, the advent of large-scale cancer genome sequencing, while initially bolstering the search for drivers, has also revealed significant challenges and paradoxes that question the sufficiency of the purely genetic, driver-focused paradigm[1,8]. A growing body of evidence highlights inconsistencies: many tumors, despite exhibiting characteristic malignant phenotypes, lack differentiable, recurrent driver mutations according to current definitions and detection methods[9]. Conversely, canonical driver mutations, such as those in *TP53* or *KRAS*, are frequently detected in apparently normal, healthy tissues, particularly with aging, without progressing to cancer[9]. Furthermore, the therapeutic applications targeting presumed driver genes, while initially effective in some cases, are often hampered by the rapid development of resistance[6,10]. Tumors can evolve mechanisms to bypass the inhibited drivers, suggesting that their maintenance relies on a broader network of cellular functions beyond the initiating oncogenes[8,11]. The sheer volume of somatic mutations within a typical tumor – often numbering in the hundreds or thousands – makes the reliable distinction between functional drivers and neutral passenger mutations conceptually and computationally challenging[1,9], and genomic profiles alone often fail to predict functional importance or therapeutic susceptibility[1,12].

These accumulating observations suggest that the limitations of the CDG paradigm are not merely technical shortcomings of driver identification algorithms, but may reflect a more fundamental incompleteness in viewing cancer solely through the lens of initiating genetic events[1]. While initiation is undoubtedly crucial, the progression and persistence of established tumors likely depend on a distinct set of molecular interactions and functional dependencies acquired during tumorigenesis that are essential for *maintaining* the complex malignant phenotype[13]. This calls for a paradigm shift towards understanding cellular states and appreciating collective or network properties that sustain cancer cell viability and proliferation[13,14]. Such a shift requires moving from a purely individual gene-centric and mutation-based view to one that embraces the cancer cell as a complex, adaptive network system of genes with acquired vulnerabilities critical for its continued existence[13].

In this context, we propose the concept of *Cancer Keeper Genes* (CKGs) that are defined as genes whose functions are indispensable for the maintenance and survival of established cancerous states of the cell, regardless of whether they were involved in initiation or harbor somatic mutations themselves[15,16]. We aim to delineate the evolution from the established malignancy initiation-focused CDG framework towards this cancer state maintenance-centric CKG perspective. We will explore how concepts like non-oncogene addiction[13], coupled with advances in systems biology, network analysis[17], and functional genomics[18], to provide the conceptual basis for understanding CKGs. We will then discuss how these concepts can be materialized into enabling tools by exploiting the theory of network controllability[14,19,20]. Ultimately, we argue that targeting these fundamental maintenance dependencies holds significant promise for developing more effective and durable cancer therapies.

## 2. Cancer Driver Genes: A Mutagenesis-based Paradigm

The current driver-gene framework in precision oncology is underpinned by two mechanistic pillars: (i) somatic mutagenesis, which generates oncogenic and tumor-suppressive driver alterations[1–3,5,21,22], and (ii) the resulting oncogene addiction of tumor cells to a subset of these aberrant signaling axes[7,23]. CDGs are operationally defined as genes whose somatic alterations confer a selective growth or survival advantage to the host cell, thereby promoting clonal expansion and malignant transformation. These alterations – collectively termed driver mutations – arise from a spectrum of mutagenic processes, including exposure to environmental carcinogens, endogenous oxidative stresses, replication infidelity, and large-scale chromosomal rearrangements. In contrast to biologically neutral passenger mutations, driver mutations rewire pivotal signaling and cell-cycle control pathways, potentially activating proto-oncogenes through gain-of-function lesions or inactivating tumor-suppressor genes through loss-of-function events. Once such an aberrant pathway is established, tumor cells frequently acquire an exquisite dependence on it – a phenomenon known as oncogene addiction[7]. Therapeutic disruption of this addiction, either by pharmacologically inhibiting the mutant oncoprotein or by functionally restoring suppressed pathways, can trigger tumor regression or apoptosis, thereby providing a conceptual and practical foundation for precision oncology[13].

## 2.1 Cancer driver gene

Cancer is fundamentally a genetic disease driven by somatic mutations in key genes that confer a growth advantage to cells[3]. Among the tens of thousands of mutations in a typical tumor, only a small subset are "driver" mutations that promote tumorigenesis, while the vast majority are neutral "passenger" mutations[2]. Driver mutations, by definition, endow the cell with a selective growth or survival advantage in its microenvironment[2]. The genes harboring such mutations are referred to as *cancer driver genes* (CDGs). To date, large-scale cancer genome sequencing efforts have identified on the order of a few hundred CDGs[1,2,5]. Early studies estimated around ~140 driver genes whose mutations can trigger tumor formation[24]. More recent pan-cancer genome analyses across thousands of tumors have expanded this catalog; for example, Bailey *et al.* (2018) identified *299* putative driver genes across 33 cancer types[5]. Each individual tumor typically contains only 2–8 driver gene mutations, which cooperate to impart the malignant phenotype, with all other mutations being passengers that do not confer a growth advantage[25]. These driver genes commonly fall into a limited number of signaling pathways regulating core cellular processes such as cell fate determination, cell survival, and genome integrity and stability[3]. For instance, many drivers affect cell cycle checkpoints, apoptosis, or DNA repair mechanisms, consistent with the known hallmarks of cancer (e.g. evading growth suppressors, resisting cell death, and sustaining proliferative signaling)[3].

CDGs can be broadly categorized as either proto-oncogenes or tumor suppressor genes, reflecting the two major genetic mechanisms of tumorigenesis[3]. Oncogenes are normal genes that become carcinogenic when mutated or overexpressed; such gain-of-function driver mutations typically confer a dominant growth-promoting effect[26,27]. Classic examples include *KRAS*, *EGFR*, and *MYC*, which often undergo activating point mutations or amplification in cancers[2]. In contrast, tumor suppressor genes (TSGs) normally restrain cell proliferation or survival, and cancer-associated mutations in TSGs are loss-of-function events (e.g., deletions and truncating mutations) that disable those growth-inhibitory functions[26]. Well-known TSG drivers include *TP53* (mutated in ~50% of cancers), *RB1*, and *PTEN*[28]. Inactivating mutations of TSGs usually follow the "two-hit" paradigm (biallelic loss) to abolish gene function[26]. Both classes of drivers ultimately deregulate the same or similar pathways – for example, an oncogenic mutation in *CDK4* (a cell-cycle activator) or a loss of *RB1* (a cell-cycle inhibitor) each leads to unchecked cell cycle progression. Thus, diverse genetic lesions converge on common biological processes[3].

Identification of driver genes in cancer genomics studies relies on distinguishing signals of positive selection from the background noise of passenger mutations[29]. Recurrent mutations found at high frequency, at functionally critical sites, or in tumors are strong candidates for drivers. Statistical algorithms identify genes with a higher mutation rate than expected by chance across large cohorts[29]. Complementary approaches search for mutational patterns characteristic of oncogenes versus tumor suppressors (enrichment of truncating mutations), as formalized in the "20/20 rule" and its variants[2]. Such integrative analyses, including those by The Cancer Genome Atlas (TCGA) and the COSMIC Cancer Gene Census, have continually refined the list of driver genes[5,30–32]. Importantly, driver mutations tend to affect evolutionarily conserved residues and impart functional changes that confer a fitness advantage to the cell[33]. In contrast, passenger mutations accumulate due to genome instability but do not contribute to cancer progression.

Recognition of CDGs has critical clinical implications. Drivers often serve as diagnostic biomarkers and therapeutic targets in precision oncology[34]. For example, the presence of an *EGFR* exon 19 deletion in a lung tumor indicates that the tumor is driven by EGFR signaling and predicts sensitivity to EGFR inhibitor drugs[35]. Indeed, a major goal of cataloging driver genes is to enable targeted therapies that specifically inhibit oncogenic drivers. Cancers "addicted" to a given driver gene are expected to regress when that gene's activity is blocked. The concept of tumors being dependent on driver genes is discussed in the next section on *oncogene addiction*. Here, the key point is that the discovery of driver genes has provided a foundation for rational drug design, exemplified by therapies like BCR-ABL tyrosine kinase inhibitors for *BCR-ABL*-driven chronic myeloid leukemia, or PARP inhibitors for BRCA1/2-mutant tumors (targeting a DNA repair vulnerability)[36,37]. As of today, dozens of driver-directed drugs have entered the clinic, and ongoing genome profiling efforts continue to unveil new driver genes and mutations that could be exploited therapeutically[31,32].

**2.2 Oncogene addiction (OA): The Achilles' heel of CDG-driven cancers**

Despite the genetic complexity of cancer, many tumors display an unexpected dependency on a single dominant oncogenic driver for their growth and survival. This phenomenon is termed *oncogene addiction* (OA) and refers to the reliance of cancer cells on continuous signaling from one overactive oncogene[7]. In other words, even though cancer cells harbor multiple mutations, inactivation of one key driver can be sufficient to markedly impair tumor viability. This concept provided a compelling rationale for molecular targeted therapies: if a cancer is "addicted" to an oncogene, then a drug that specifically inhibits that oncogene should have potent and selective antitumor effects like "removing the tumor's Achilles' heel"[7]. Over the past two decades, extensive preclinical and clinical evidence has validated OA as a real and exploitable vulnerability in many types of cancer[38,39].

A quintessential example of OA is chronic myelogenous leukemia (CML), which is dependent on the *BCR-ABL* fusion oncogene[40] The constitutively active BCR-ABL tyrosine kinase drives CML, and the introduction of the BCR-ABL inhibitor imatinib has demonstrated dramatic clinical responses[41]. Patients treated with imatinib have shown rapid hematologic remissions, providing the first proof-of-concept that shutting off a single oncogenic driver can effectively control a malignancy. Imatinib's success transformed CML from a fatal disease into a manageable chronic condition; the 5-year survival rate improved from ~30% in the pre-imatinib era to over 85–90% with imatinib therapy[36]. CML has been dubbed the "poster child" of OA, as most CML cells undergo apoptosis when BCR-ABL signaling is ablated. Similarly, in gastrointestinal stromal tumors (GIST), activating mutations in *KIT* drive the cancer, and these tumors respond markedly to KIT inhibitor therapy[42]. These cases illustrate how oncogene-addicted cancer is highly sensitive to a targeted agent.

OA is not limited to hematologic malignancies. Many solid tumors also exhibit this behavior. For instance, a subset of non-small cell lung cancers (NSCLC) is driven by a mutant EGFR; treating these patients with EGFR tyrosine kinase inhibitors (TKIs) yields response rates of approximately 70% and significantly prolongs progression-free survival compared to chemotherapy[43]. Another example is *BRAF*-mutant melanoma, where BRAF inhibitors produce tumor regressions in ~50% of patients and improve outcomes relative to prior standard therapies[44]. In ALK-rearranged lung cancer, ALK inhibitors induce high response rates (around 60%) and durable control in oncogene-addicted tumors[45]. These clinical experiences reinforce a general theme: targeting an essential oncogenic driver often leads to a "marked improvement in initial patient responses" when directly compared with conventional therapy. In many trials, oncogene-targeted agents have achieved higher response rates and longer disease control than non-targeted treatments, validating the strategy of hitting the tumor's dominant growth engine[46].

Mechanistically, why do cancer cells become so singularly dependent on one oncogene? One hypothesis is that during tumor evolution, the activated oncogene orchestrates a network of pro-survival and proliferation signals, while cancer cells concurrently adapt to tolerate the oncogene's aberrant signaling. This brittle "balance" can render the cancer cell highly vulnerable to the acute withdrawal of the oncogenic signal – a concept termed "oncogenic shock", wherein turning off the oncogene triggers apoptosis before the cell can compensate[47]. Additionally, oncogene often provides necessary signals that cancer cells cannot easily replace; thus, its inhibition creates a lethal cellular crisis. Whatever the exact mechanism, the practical outcome is that many tumors exhibit an all-or-none dependence on key oncogenes[48].

An important corollary of OA is the challenge of acquired resistance. Although initial responses to targeted inhibitors can be dramatic, they are often followed by tumor relapses due to the development of resistance mechanisms. Common resistance mechanisms include secondary mutations in the oncogene that prevent drug binding (e.g., the EGFR T790M or BCR-ABL T315I mutations), amplification or overexpression of the oncogene, activation of parallel signaling pathways that bypass the blocked oncogene, or histological transformation of the tumor cells[49]. These changes essentially restore the critical signaling that the tumor is addicted to, or activate an alternative driver, thereby undermining the effects of the drug. The recurring observation that resistant cancer cells often reactivate the same pathway underscores that the tumor was truly addicted to that pathway[50]. For example, most EGFR-mutant lung cancers that become resistant to first-generation TKIs acquire an EGFR gatekeeper mutation (T790M) or bypass through MET amplification, both of which re-engage the EGFR/ERBB signaling axis[51]. This persistence of dependency, even in the face of drug pressure, "makes the most compelling argument for oncogene addiction" in human cancers[50]. New generations of inhibitors and combination therapies are being developed to overcome resistance and induce more durable remissions. Nevertheless, the principle of OA has firmly established a foundation for precision oncology – if an oncogenic driver can be identified and effectively targeted, patient outcomes can be significantly improved.

It is worth noting that not all cancers have an obvious druggable oncogene to target. Some oncogenes, like *RAS*, have historically been difficult to inhibit pharmacologically[52]. Some tumors are driven by loss-of-function mutations in tumor suppressors (e.g., *TP53* and *RB1*). Additionally, tumors with multiple co-driving events may not be solely dependent on one gene[53]. In such cases, targeting a single oncogene may be insufficient or inapplicable. These observations prompted an exploration of whether cancers have alternative dependencies beyond the classic oncogenes, leading to the concept of *non-oncogene addiction*.

**3. Beyond Cancer Driver: Non-oncogenic Dependencies and Cancer Dependency Map**

While CDG and OA primarily focus on the initiation and progression of oncogenesis driven by mutations in a limited number of oncogenes or tumor suppressors, it is equally, if not more, clinically relevant to examine how cancer cells maintain the homeostasis of malignancies. Disrupting such homeostasis presents an opportunity to induce apoptosis by targeting the uncontrolled cell proliferation. However, many genes work together to maintain this balance and may not necessarily be mutated. Rather, they function collaboratively within complex regulatory networks. As such, pinpointing the specific genes crucial for maintaining malignant homeostasis becomes a significant challenge for cancer therapy.

**3.1 Non-oncogene addiction (NOA): Dependence on non-mutated but essential genes**

As tumors evolve, cancer cells must adapt to various intrinsic stresses, dynamic microenvironments, and hostile exogenous environments. To survive and proliferate despite these challenges, cancer cells become dependent on numerous genes and pathways that are not oncogenic drivers *per se* but are nonetheless crucial for cell survival. This phenomenon has been termed *non-oncogene addiction* (NOA), referring to the cancer's "addiction" to normal, non-oncogenic cellular functions that help cope with the stresses of oncogenic transformation[13]. In essence, cancer cells leverage a variety of support pathways – often the cellular stress response, DNA damage repair, metabolic regulators, etc. – to buffer the detrimental effects of having a dysregulated, fast-growing, genomically unstable state[54]. While these non-oncogene dependencies do not *initiate* cancer, they are essential for maintaining the cancerous state and disabling them can selectively kill cancer cells without affecting normal cells as severely.The concept of NOA[11,13] extended the classic *Hallmarks of Cancer*[3] to include the stress phenotypes inherent to cancer cells. Cancer cells experience elevated proteotoxic stress (resulting from the accumulation of misfolded proteins), metabolic stress, replicative stress (from rapid cell division and DNA replication), oxidative stress (due to high levels of reactive oxygen species), and DNA damage stress, among other stressors[55]. To survive these pressures, tumors become reliant on stress-support pathways[56]. For example, cancer cells often upregulate chaperones and heat-shock proteins to manage proteotoxic stress. *Heat shock factor 1* (HSF1) is a transcription factor that induces the expression of heat-shock proteins, such as HSP70 and HSP90/HSP90, in response to protein-damaging stress[57]. HSF1 itself is rarely mutated in cancer and does not cause transformation; however, tumor cells frequently require HSF1 activity to sustain their malignant growth[57]. In mouse models, HSF1 knockout impairs tumor formation, yet HSF1 is largely dispensable in normal cells under non-stressful conditions[58]. This indicates that cancer cells are addicted to HSF1 and the proteostasis network it controls – a clear example of NOA supporting the tumor's proteome stability.

Likewise, cancer cells endure oxidative stress due to oncogenic metabolism and dysfunctional mitochondria, and thus, they become dependent on robust antioxidant systems. *NRF2* (NFE2L2) is a transcription factor that upregulates antioxidant genes, e.g., glutathione synthesis enzymes and superoxide dismutases. Many tumors exhibit constitutive activation of the NRF2 pathway or related antioxidant programs to neutralize excess reactive oxygen species (ROS). While NRF2 activation is not the original driver mutation in most cases, the functional reliance on NRF2 for redox homeostasis represents a non-oncogene dependency[59,60]. Targeting molecules like NRF2 or modulating redox balance can therefore harm cancer cells that have high oxidative stress levels but spare normal cells with lower basal stress.

Another crucial stress phenotype in cancer is DNA replication stress and genomic instability. Oncogene activation (e.g., MYC overexpression) pushes cells into rapid S-phase cycles and can cause replication fork stalling and DNA breaks[61]. Tumors adapt by relying on the cell's DNA damage response (DDR) pathways to repair DNA and manage stalled forks. Ataxia Telangiectasia and Rad3-related (*ATR*) kinase and its effector, Checkpoint kinase 1 (CHK1), are central players in the replication stress response, pausing the cell cycle and stabilizing forks when DNA is damaged or replication is perturbed[62]. Cancer cells, especially those deficient in other checkpoint controls, such as p53, often become highly dependent on ATR/CHK1 to survive their continuous replication stress[63]. Indeed, experimental studies have shown that MYC-driven cancers are selectively sensitive to ATR or CHK1 inhibitors, which induce catastrophic DNA damage in these cells while sparing normal cells with intact checkpoints[64]. This synthetic-lethal interaction (oncogene-induced stress requiring a non-oncogene DDR factor) is being explored therapeutically: ATR inhibitors are in trials, aiming to exploit the addiction of certain tumors to ATR-mediated repair.

NOA overlaps conceptually with the strategy of synthetic lethality, wherein the combination of a tumor's mutation and a targeted inhibition leads to cell death[65]. A prominent example is the use of PARP inhibitors in BRCA-mutant cancers[37]. *BRCA1* and *BRCA2* are tumor suppressors involved in homologous recombination DNA repair; their loss is a driver of breast and ovarian cancers[66]. Those cancer cells become extremely reliant on the PARP-mediated single-strand break repair pathway as a backup. PARP itself is not an oncogene – it is required for DNA repair in all cells – but BRCA-deficient tumor cells are *uniquely* sensitive to PARP enzyme inhibition because they cannot repair DNA by either pathway, leading to tumor cell death while normal BRCA-proficient cells withstand the treatment[67]. This synthetic lethal approach exemplifies NOA: the BRCA-null cancer is "addicted" to PARP for survival. In clinical practice, PARP inhibitors have achieved significant success in BRCA-mutated cancers[68], validating the idea that targeting a non-oncogenic dependency can yield tumor-selective effects.

Because NOA arises from the unique stresses of cancer, it presents a rich landscape of potential therapeutic targets that are distinct from classic oncogenes. Notably, targeting non-oncogene dependencies may broaden the scope of treatable cancers, including those without an obvious druggable driver. For instance, even if a cancer is driven by *KRAS* (an oncogene that is tough to inhibit directly), it may have exploitable non-oncogene dependencies, such as a need for heightened autophagy or for certain metabolic enzymes. This is because KRAS-driven cancers often rewire their metabolism and can become dependent on glutaminase[69]. Inhibition of such non-oncogenic targets can cripple the cancer cell's adaptive machinery. This approach has been termed "drugging the cancer fitness genes," as opposed to targeting traditional oncogene pathways[70]. Many drug development efforts have been pursuing agents against these non-oncogene targets, such as HSP90 inhibitors to target chaperone addiction, glutaminase inhibitors to exploit cancer metabolic dependencies, and checkpoint kinase inhibitors to induce lethal DNA damage in genomically unstable tumors[71].

It is crucial that non-oncogene targets are selected so that cancer cells are significantly more vulnerable to their loss than normal cells. The therapeutic window arises because normal cells, not experiencing oncogenic stress at extreme levels, can survive without these stress-coping mechanisms, whereas cancer cells cannot. For example, as noted, HSF1 or CHK1 knockout mice are viable and largely normal[72]; yet, tumors are profoundly impaired without those factors[57]. This differential requirement gives hope that drugs against non-oncogene dependencies will be selective for tumors. However, toxicity can still be a concern if normal tissues that have high proliferation (e.g., bone marrow and gut epithelium) are affected[73]. The challenge is to find targets that maximize tumor selectivity.

In short, the NOA concept has expanded the scope of cancer drivers and cancer therapy beyond mutant oncoproteins. It recognizes that the context of cancer (oncogenic stress and rewired homeostasis) also creates "Achilles' heels" in pathways that are not themselves oncogenic. Combining therapies that hit both the oncogene (if one exists) and key non-oncogene dependencies may yield synergistic effects, driving cancer cells into an intolerable state of stress overload. This approach, sometimes referred to as *stress sensitization*, may help overcome resistance and achieve more profound and durable tumor remissions[54].

**3.2. Cancer dependency map: A genome-wide empirical approach**

Although the concept of NOA has demonstrated efficacy in certain cancer contexts[11,54], it remains a conceptual framework lacking a standardized methodology for identifying genes that disrupt cellular homeostasis and serve as viable drug targets. A major obstacle in pinpointing effective non-oncogene targets lies in the intricate interactions, correlations, and dependencies among genes that function in the same cancer-related pathways or parallel pathways that may compensate for one another's loss[14,74]. This functional redundancy often conceals potential therapeutic vulnerabilities.

Classical genetic studies based on the theory of Synthetic Lethality (SL) have provided valuable insights into genetic interactions whereby the simultaneous perturbation of two genes results in cell death, whereas disruption of either gene alone is tolerated[70,71]. Such studies have been instrumental in identifying gene pairs that can be exploited therapeutically, exemplified by the use of PARP inhibitors in BRCA-mutated cancers[75,76]. However, traditional SL approaches are limited by scalability issues, restricting their applicability across the genome and diverse cancer types[77].

To fully realize the potential of Oncogene Addiction (OA)[11,78] and NOA[11,13] on a genome scale, it is imperative to conceptualize cancer cells as complex networks of interacting genes to facilitate identifying cancer-associated dependencies across heterogeneous disease contexts[14,17]. The Cancer Dependency Map (DepMap) project was thus initiated as a large-scale collaborative effort aimed at systematically charting the genetic vulnerabilities of cancer cells[79,80]. Its objective is to determine which genes are essential for the survival and proliferation of cancer cells across hundreds of human tumor cell lines representing a broad spectrum of cancer types.

DepMap integrates high-throughput functional genomics with computational modeling to generate a comprehensive, cell–line–specific compendium of genetic dependencies[81]. The project employs state-of-the-art genome-scale experimental techniques, including CRISPR-Cas9 knockout screens, which enable precise gene disruption[18,82]; RNA interference (RNAi) screens, which transiently reduce gene expression[83]; and small-molecule perturbation assays, which evaluate the effects of chemical inhibitors on gene-regulated pathways[80,84,85]. These methodologies systematically perturb gene regulatory and metabolic networks, allowing assessment of the consequences of silencing nearly every gene in the genome on cancer cell viability.

Importantly, DepMap combines such functional genomic data with multi-omics datasets, including cancer genomics, transcriptomics, proteomics, and detailed cell line characterization[81], to distinguish between genes that are broadly essential for cell survival and those that represent context-specific dependencies. The latter category comprises genes whose loss is lethal only in particular cancer types or genetic backgrounds, thereby representing ideal therapeutic targets that can selectively eradicate tumor cells while sparing normal tissues[86].

The DepMap initiative constitutes a significant advance in cancer research by providing a scalable and systematic platform for the discovery of both canonical cancer driver genes[2,87] and genes critical for maintaining the malignant phenotype. For instance, it has facilitated the identification of genes uniquely essential in tumor cells harboring mutations in key oncogenes or tumor suppressors such as KRAS, TP53, and BRCA1[88]. This has enabled the uncovering of synthetic lethal interactions that can be therapeutically exploited, wherein inhibition of a gene is selectively toxic in the presence of a specific oncogenic mutation[79,89].

In short, the DepMap offers a transformative, function-first perspective on cancer biology. By shifting the focus from static genomic alterations to dynamic gene essentiality profiles[14,90], it equips researchers and clinicians with a powerful resource to, on a genome-wide scale, identify, prioritize, and therapeutically target the genes upon which tumors are truly dependent for survival and growth[76,91,92].

**4. Cancer Keeper Genes: A Paradigm Centered on Malignancy Maintenance**

In retrospect, over the past few decades, since the inception of the concept of cancer driver genes, cancer research has shifted its major focus from an individual gene-centric perspective, which primarily focused on cancer driver genes, to a genome-scale and network-oriented viewpoint, culminating in the latest effort by the DepMap consortium[80,84,85]. The DepMap's system-scale exploration of cancer gene dependencies across diverse cell lines has revealed co-dependent and synthetic lethal interactions among multiple genes[93,94], highlighting complex functional relationships and potential combinatorial vulnerabilities in cancer.

This technical migration from individual genes to gene clusters also marks a conceptual shift from cancer driver genes, pillared by the OA concept, to Cancer Keeper Genes (CKGs), i.e., genes that maintain malignancy homeostasis[15]. Inspired by the network medicine perspective[14], the concept and computational method for CKGs are complementary to the empirical approach of the DepMap. The CKG approach rests on network control theories, particularly the network structural controllability[19,20,95]. Interestingly, this theory was first applied to identify cancer driver genes in regulatory networks, discussed next.

**4.1. Network driver nodes as candidate cancer driver genes – A network controllability approach**

The gene regulatory network (GRN) of a cancer cell constitutes a nonlinear dynamical system whose topology and interaction strengths vary continuously throughout the tumor life cycle. Attractor-landscape analyses reveal that oncogenesis corresponds to trajectories that escape physiological basins and settle in pathological ones, driven by temporally shifting genetic and epigenetic perturbations[90]. Stage-resolved and single-cell studies reveal pervasive edge rewiring and reversible switch-like configurations, indicating that therapeutic leverage resides in modulating dynamic connectivity rather than targeting static hubs[96,97]. To translate these insights into precise interventions, cancer research must incorporate network-dynamical modeling and control theory, which together can quantify how transient or sustained perturbations propagate through evolving networks and identify the nodes or edges that yield the greatest influence over system-level trajectories.

Network control theory provides a mathematically rigorous vocabulary for this task. Building on the structural controllability theorem, Liu et al. demonstrated that any linear time-varying network can, in principle, be driven from an arbitrary initial state to any desired final state by externally manipulating a minimal set of driver nodes[19]. Although biological GRNs are nonlinear, subsequent work has shown that structural criteria derived from a linear approximation can yield informative control points for a broad class of biological networks, including neuronal circuits, metabolic pathways, and protein-interaction maps[95,98]. Computational implementations typically construct a directed, weighted graph representing transcription-factor binding, signaling, or protein interactions, after which algorithms such as maximum-matching or integer-linear-programming enumerations produce one or more minimal driver-node sets[14].

The application of this viewpoint to cancer has been shown to be valuable for identifying potential CDGs. Vinayagam et al. mapped the human protein–protein interaction network onto a controllability framework and recovered a large proportion of experimentally validated oncogenes and clinically actionable drug targets, thereby corroborating the utility of driver-node analysis for uncovering functionally indispensable genes[98]. More recently, Guo et al. incorporated patient-specific genomic and transcriptomic data to derive individualized GRNs; their personalized controllability analysis revealed unique driver nodes for each patient, many of which displayed mutational or expression patterns consistent with aggressive disease and poor prognosis[99]. These studies collectively suggest that network-level controllability, rather than mutation frequency alone, offers a more mechanistic criterion for defining genes indispensable to malignant states[17,95].

Nevertheless, the current driver-node methodology faces several theoretical and practical limitations. First, large GRNs admit an astronomically large number of alternative control configurations; exhaustive search of all these configurations is #P-hard, and heuristic approximations are unlikely to guarantee biologically optimal solutions[100]. Second, structural controllability prescribes the simultaneous manipulation of all nodes in a designated driver set, whereas current genome-editing technologies cannot induce coordinated, instantaneous perturbations across dozens of loci in vivo[101]. Third, standard algorithms assume *a priori* knowledge of which regions of the global state space correspond to "normal" or "cancer" phenotypes, yet this mapping is rarely available with sufficient resolution[96]. Driver nodes derived solely from topological information may therefore lack functional relevance if they do not coincide with genes that modulate the biochemical parameters separating the competing attractors. These challenges highlight the need for revised controllability formalisms.

**4.2. Cancer keeper genes – A total network controllability approach**

Cancer driver genes have traditionally emphasized the mutational events that trigger malignant transformation, that is, those genetic alterations capable of converting a normal cell into a cancer cell[2]. Merely identifying these "ignition" genes, however, is insufficient to elucidate how a malignant phenotype is stably preserved once transformation is complete[13]. To address this limitation, we recently introduced the concept of Cancer Keeper Genes (CKGs), which designate genes indispensable for sustaining the malignant state of cancer cells[15]. In contrast to driver genes, CKGs do not need to be mutated in normal tissue and do not directly initiate tumorigenesis; rather, they provide the functional capacity that facilitates established tumor cells to maintain the homeostasis of malignant cells, resulting in aberrant proliferation and survival[11,13]. If driver genes can be likened to the initial "spark" that ignites tumor formation, CKGs constitute the "fuel supply" that keeps the malignancy burning. Systematic identification of these core maintenance genes reveals the network-level dependencies of tumor cells on particular regulatory pathways, thereby offering new avenues for therapeutic intervention[79].

To detect such indispensable network nodes, we introduced the concept of control hubs[102]. Specifically, a node is deemed a control hub when it is traversed in every feasible control path of the network – most notably when it resides in an intermediate, rather than initial or terminal, position along these paths[20]. Any attempt to steer the global system state must therefore pass through this hub. The analogy is that of a critical intersection in a transportation network: once blocked or disrupted, network-wide traffic becomes unsustainable. Within the gene-regulatory circuitry of tumor cells, these control hubs furnish the structural scaffolding that keeps the system locked in the "cancer attractor," and they thus correspond to Cancer Keeper Genes.

Network analyses across multiple tumor types have identified substantial numbers of CKGs[15]. In bladder cancer (BLCA), for example, analysis of a network comprising 7,030 genes and 103,360 regulatory interactions identified 660 control hubs[15]. Although numerically modest, these CKGs exhibit high connectivity and occupy key positions in the regulatory network. Functional enrichment analysis revealed a marked concentration of genes in various cancer-related pathways, including cell signaling, cell cycle regulation, transcriptional control, and protein translation modification, whereas no comparable enrichment was observed for basal metabolic or structural components. These distinctions indicate a predominant role for CKGs in governing cell-state maintenance and fate determination, consistent with their non-housekeeping characteristics in tumors. Notably, CKGs tend to be evolutionarily conserved and essential for human cells, underscoring their pivotal functions in both human evolution and cellular physiology, which resonates with their centrality in tumor maintenance.

To distinguish the most druggable targets among CKGs, we further defined "sensitive control hubs," i.e., control hubs that are exceedingly vulnerable to minor structural perturbations of the network and readily lose their hub status when even a few edges are altered[15]. The network's reliance on these nodes is thus extreme, and slight interference can compromise overall controllability, destabilizing the tumor attractor. Among the 660 CKGs in the bladder cancer network, 35 qualify as sensitive control hubs, and almost all cluster within two canonical oncogenic pathways: the cell cycle pathway and the p53 pathway. The former encompasses core regulators, such as E2F transcription factors, the CDK family, and Cyclin family members, whereas the latter includes p53-related genes pivotal for growth arrest and apoptosis. These findings suggest that, despite the diverse molecular adaptations acquired during tumor evolution, the capacity for limitless proliferation and evasion of apoptosis remains a non-negotiable requirement; reactivation or inhibition of key genes along these two pathways can neutralize the malignant growth potential[91]. Similar patterns have been observed in preliminary analyses of other cancers – including cervical carcinoma and head-and-neck squamous-cell carcinoma – corroborate a shared evolutionary dependency on these core pathways (data unpublished). The results align with the established tenet of tumor biology, which posits that aberrant control of the cell cycle and disrupted apoptosis are essential for malignant transformation. This finding also provides network-level evidence that therapeutic interference with these two pathways may yield broadly effective anticancer strategies.

**4.3. Cancer keeper genes – A unified concept and analytical framework for malignancy maintenance**

The introduction and characterization of CKGs have put forward a unified theoretical framework centered on maintaining the homeostasis of malignant cells. It unifies the concepts of NOA and DepMap, meanwhile explicitly separating tumor persistence from tumor initiation.

In this unified framework, a tumor is construed as a systems-level state that is triggered by driver mutations and thereafter upheld by reconfiguring the network of gene interactions and regulations[90]. Driver mutations propel a normal cell into an aberrant attractor – effectively "flipping the cancer switch". Once malignant cells have arisen, preservation of their pathological steady state demands a continuous input from a limited set of critical network-control nodes that provide a "sustained power supply"[11]. These nodes constitute the CKG set and encompass 1) principal mutated drivers, insofar as they remain mechanistically operative, together with 2) indispensable, non-mutated functional supporters. Dependence on CKGs can manifest as oncogene addiction when a CKG is itself a driver gene, as non-oncogene addiction when a CKG acts in a supportive capacity, or as a combination of both modalities[78].

Consequently, the CKG framework integrates the mutation-centered viewpoint of classical cancer genetics with the network-control perspective of systems biology: it recognizes that tumors originate in discrete molecular events such as gene mutations, while simultaneously emphasizing that the malignant phenotype represents a global network-dynamic phenomenon that necessitates system-level keystone nodes for its maintenance[103].

From the perspective of OA, the primacy of mutated driver genes is underscored; yet CKGs encompass, but are not limited to, driver genes[14]. Any gene on which tumor cells are acutely dependent, irrespective of whether it harbors driver mutations, can be considered a "maintenance" gene for the cancer state. For instance, a tumor persistently reliant on mutated EGFR signaling can be said to exhibit OA, making *EGFR* both a driver gene and a CKG[104]. However, CKGs operate under a broader sense, not requiring a gene itself to be implicated in the initiation of tumorigenesis. Thus, even if a gene has never undergone mutation, it may be categorized as a CKG as long as it plays a critical role in the survival of tumor cells[13]. In effect, oncogene addiction is a subset of CKG dependence (i.e., CKGs rooted in driver mutations), while the more inclusive CKG framework extends to key genes without direct oncogenic mutations. Both approaches share a common objective in identifying major single-gene vulnerabilities in tumor cells. They simply differ in their scope – one focuses on driver mutations, whereas the other encompasses all genes essential to cancer maintenance[11].

CKGs are also closely aligned with NOA. Most CKGs are expected to be part of the NOA category, meaning they lack cancer-causing mutations, yet remain indispensable for preserving the tumor's malignant phenotype[13]. NOA provides a rich roster of potential genes (e.g., DNA repair and stress responses) for CKG identification, whereas the CKG concept adds a more rigorous network- and control-theoretic lens. Effectively, CKGs represent a formal extension of NOA, which was developed from empirical observations. While NOA focuses on whether a gene is mutated or not, CKGs concentrate on functional necessity. Both, however, share the premise that cancer cells may be "addicted" to certain genes outside the typical circle of drivers[105]. Many stress-response genes, such as CHK1, ATR, HSP90, and BCL2, align with the concept of NOA and tend to be highlighted by the CKG approach for maintaining tumor homeostasis[63,71]. By focusing on genes that sustain the tumor's current state, CKGs effectively represent a system-level application of NOA[11].

Furthermore, the CKG concept is intimately tied to that of network hub genes. CKGs commonly serve as central control nodes or hubs within the network, orchestrating downstream pathways to preserve the tumor's stability[14]. For example, in studies on bladder cancer, essential genes in both the cell cycle and p53 pathways have been identified as CKGs, underscoring their essential, hub-like role in controlling cell proliferation and death. While network hub genes are defined primarily based on topological features, CKGs emphasize functional indispensability under cancer conditions[101]. Not all hubs are necessarily required by tumor cells; some may possess redundant pathways and do not constitute genuine weak points in the network[106]. In contrast, CKG selection pinpoints genes that lack viable alternatives for maintaining the malignant network state[107]. One may thus regard CKGs as functional hub genes, located not only centrally in the network's structure but also essential to stabilizing its dynamic cancer-driven behavior[14]. Network hub analyses facilitate discovering potential CKGs, and the CKG framework refines this method through control theory, identifying those nodes that sustain the overall tumor state[98]. In this reciprocal relationship, higher topological centrality increases the likelihood of a gene being a CKG, while CKG analysis helps deepen understanding of which nodes truly matter in the cancer network[103].

Finally, CKGs are inextricably linked to and extend synthetic lethality strategies. Becoming a CKG implies that a gene exhibits lethal dependency in certain tumor genotypes but not in normal cells, essentially a broad form of synthetic lethality (the cancer state versus the normal state)[91]. In other words, CKGs often display an essentiality unique to the cancer context[76]. One pertinent example is PARP1, which is indispensable in BRCA-deficient tumors[37]. In that setting, PARP1 can be viewed as a CKG whose inhibition proves devastating to the malignant state[76]. While the CKG concept does not explicitly stipulate the presence of a partner gene, the other factor is implicitly the aberrant network configuration of the cancer itself[91]. Thus, identifying CKGs essentially locates the special dependencies that differentiate tumor cells from their normal counterparts, mirroring the goal of synthetic lethality in discovering lethal combinations[108]. However, while conventional synthetic lethality often relies on known driver mutations, such as BRCA deficiency, to identify potential partners, the CKG strategy can more broadly detect any single-gene dependency emerging under a cancer-specific network state, including cases in which multiple gene perturbations converge to create such a reliance. Consequently, the CKG framework enriches the pool of candidate targets for synthetic lethality, and synthetic lethality in turn offers a roadmap for validating these functional dependencies and guiding drug development. Targeting a CKG should selectively destabilize tumor networks, while normal cells remain more resilient due to the presence of redundant or compensatory pathways[107]. Such an approach offers new therapeutic opportunities. For instance, if a particular CKG is crucial in KRAS-mutant tumors but dispensable in KRAS wild-type cells, blocking it would constitute a synthetic lethality strategy against the KRAS alteration[69]. Hence, CKGs provide potential effective therapeutic targets, and synthetic lethality offers a translational pathway for their clinical application[108].

Taken together, the CKG concept embodies a comprehensive synthesis of cancer cell vulnerabilities. It inherits key insights from earlier frameworks – namely, that cancer cells depend on special addictive genes, typically occupying central positions in cellular networks, and affording selective lethal outcomes when disrupted – yet adds a novel dimension by applying systems biology[103]. Instead of relying solely on mutation frequency or experimental observations, the CKG model integrates global network properties to assess gene importance, advancing beyond conventional methods of target identification[14]. This theoretical synthesis provides a more holistic view of tumor dependencies and opens up new horizons for discovering therapeutic targets[107].

**5. Clinical Translation: Targeting CKGs for Durable Cancer Control**

The search for CDGs has been a dominating theme in pharmacology and pharmacogenomics in recent years[21,109]. Therapies targeting such driver oncogenes have yielded dramatic tumor regressions, yet long-term success is often limited by the cancer's ability to adapt and develop resistance[110]. One fundamental challenge is tumor heterogeneity: different patients, even different cells within one tumor, harbor distinct driver mutations, and no single oncogene target is universally essential across all cancer cells[53]. Such genetic heterogeneities further advocate a paradigm shift from the genes that *start* cancer to those that *maintain* the malignant state[11,15].

**5.1 Potential for durable tumor control and resistance prevention**

A major motivation for targeting CKGs is the promise of more durable tumor control with less opportunity for resistance. Cancers are adept at escaping single-target drugs by rerouting signaling pathways, activating backups, or mutating the drug's binding sites[111,112]. These adaptive escape mechanisms reflect the underlying redundancy and plasticity of cancer signaling networks – if one pathway is blocked, an alternative can often compensate[74]. By aiming at CKGs, therapy strikes at *central hubs of the network*, making it far more difficult for the tumor to compensate or evolve a quick resistance. Indeed, network analyses indicate that removing a key hub can destabilize the entire cancer network, leading to collapse of malignancy[113]. In practical terms, this could translate to prolonged remission or tumor stasis: the cancer becomes unable to easily find a detour around the blocked CKGs, resulting in sustained control.

Evidence supporting this potential is beginning to emerge from preclinical studies. For example, we recently identified 660 CKGs in bladder cancer that maintain the tumor's gene-regulatory network[15]. Strikingly, *all* members of the cell cycle regulation and p53 tumor suppressor pathways were identified as CKGs in that cancer model built with experimental data from previous studies, underscoring that fundamental proliferation and survival circuits are indispensable to malignancy. We further refined a subset of CKGs, termed as sensitive CKGs (sCKGs) whose network hub status could be easily disrupted by minor perturbations, making them especially attractive drug targets. Experimental validations using cancer cells and a mouse model showed that inhibiting just one of these sensitive hubs can have an outsized effect: six sCKGs (including examples like *FGFR3* and *EP300*) were experimentally targeted, and each intervention effectively suppressed cancer cell growth in vitro and in tumor-bearing mice. Such results illustrate the principle that hitting a keeper gene can produce a broad anti-tumor response. The reason is intuitive – a driver oncogene is like a single gas pedal stuck down, whereas a keeper gene is more akin to the engine's control unit; shutting down the latter stalls the whole malignant "machine," not just one pathway. The early findings also hint that resistance to CKG inhibition may be less frequent or slower to develop, since the tumor would need to rewire large portions of its network or overcome fundamental stress constraints to compensate for the lost function. This hypothesis aligns with observations in other contexts: tumors heavily reliant on stress-support pathways (heat shock proteins, DNA repair mechanisms, etc.) often have *limited alternative options* if those supports are removed[114]. By targeting such non-redundant support genes, therapy can induce a collapse of cancer cell fitness, potentially yielding longer-lasting remission compared to the transient responses of many current targeted drugs[115].

It is worth noting that clinical experience indirectly supports the notion that broadly targeting a tumor's dependencies leads to more durable control[116]. One illustrative example is the success of PARP inhibitors in cancers with homologous recombination deficiency (such as BRCA-mutated ovarian cancer)[37]. PARP is not an oncogenic driver, but tumor cells lacking BRCA1/2 are extremely dependent on PARP-mediated DNA repair – a classic "keeper" function. As maintenance therapy after chemotherapy, PARP inhibitors have significantly prolonged progression-free survival in advanced ovarian cancer[117], demonstrating how exploiting a cancer's intrinsic dependency can translate into durable clinical benefit. Similarly, in hormone receptor-positive breast cancer, combining endocrine therapy with a cell-cycle CKG target (CDK4/6 kinase) has become standard because it yields far more sustained tumor control than hormone therapy alone[118]. These examples underscore that when therapy attacks the cancer's maintenance mechanisms (whether via synthetic lethality, cell-cycle blockade, or other keeper functions), the outcome can be longer-lasting responses and slower emergence of resistance. CKG-based treatments generalize this principle, systematically identifying each tumor type's critical maintenance genes and drug them for maximal and enduring impact.

## 5.2 Integrating CKG-targeted therapies into treatment strategies

Translating the concept of CKGs into clinical practice will likely involve incorporating CKG-targeted agents into existing therapy regimens. CKG inhibitors may enhance and extend the efficacy of established modalities, including driver-focused targeted inhibitors, chemotherapy, or immunotherapy. The underlying rationale is that while a conventional agent debulks the tumor by inhibiting its primary growth signal, the CKG inhibitor could prevent the cancer from activating alternative survival pathways all at once. For instance, combining a BRAF inhibitor with an agent that broadly impairs cell survival signaling could thwart the pathway reactivation known to lead to melanoma relapse[111] . Early studies in breast cancer illustrate a promising approach: adding a cell-cycle inhibitor, such as a CDK4/6 blocker, to endocrine therapy not only enhances initial tumor regression but also significantly delays resistance, resulting in prolonged progression-free survival[118]. A similar rationale might apply to lung cancer, where pairing an EGFR inhibitor with a stress-response inhibitor, for example, targeting HSP90 or autophagy, could simultaneously block the oncogenic driver and the cell's protective machinery[104]. By designing combination regimens that incorporate CKG inhibitors, it becomes possible to address both the tumor's primary vulnerability and its fundamental survival mechanisms simultaneously, thereby offering deeper and more durable responses than either strategy alone[108].

CKG-targeted agents also hold promise as maintenance therapies, which are low-intensity treatments administered after the bulk of the tumor has been reduced, to prevent regrowth. This principle is already used in malignancies such as ovarian cancer and leukemia, where maintenance therapy helps control minimal residual disease based on the premise that any surviving malignant cell can eventually drive relapse[119]. CKG inhibitors could reinforce this strategy by capitalizing on the fact that residual tumor cells, left behind after surgery, chemotherapy, or targeted therapy, are often stressed, highly reliant on keeper genes, and therefore vulnerable[114]. Administering a CKG-targeted agent in this setting could eradicate these residual cells or force them into a dormant, nonthreatening state. This approach may prove particularly effective in cancers that exhibit frequent late relapse, such as certain leukemias or breast cancers, where extended maintenance therapy is already standard practice[120]. By targeting the genes that sustain core survival processes in microscopic disease, CKG-directed maintenance therapy aims to eliminate residual cell populations more definitively, thereby improving cure rates.

A further compelling application of CKG inhibition arises in salvage therapy for refractory or relapsed cancers[6]. Tumors that have progressed through multiple lines of treatment commonly exhibit high adaptive capacity, harboring a patchwork of new mutations and activated bypass pathways that render them resistant to therapies focused on a single pathway[111]. CKG-directed interventions bypass this complexity by focusing on fundamental cell-state requirements rather than any single signaling node, exploiting a shared vulnerability that persists across diverse resistant subclones[14]. For example, a heavily pretreated metastatic tumor that has become resistant to hormone therapy, PI3K inhibitors, and CDK4/6 inhibitors may remain dependent on a key mitotic checkpoint kinase or epigenetic regulator not targeted by prior therapies[121]. Introducing a CKG inhibitor in this setting could yield meaningful tumor control even when established therapies have failed. A relevant clinical scenario is EGFR-mutant lung cancer that develops resistance to tyrosine kinase inhibitors: rather than attempting to neutralize each new resistance mutation, one could administer an experimental agent that inhibits a master transcriptional regulator essential to lung cancer cells, thereby impacting the heterogeneous resistant population as a whole[111]. By reconfiguring the cancer's underlying network, such salvage therapy may resensitize tumors to other drugs or at least halt further progression[14]. In this manner, CKG-targeted strategies offer a necessary addition to the therapeutic arsenal, especially when conventional treatments have been exhausted.

**5.3 Optimizing CKG-targeted therapies**

Not every cancer or disease stage will benefit equally from a CKG-based strategy – identifying the sweet spots for translational impact is crucial[122]. CKG could be most transformative in contexts where traditional therapies underperform, such as highly heterogeneous advanced cancers and states of minimal residual disease[123]. In late-stage or metastatic cancers, tumors often harbor a diversity of genetic clones and aberrantly activated pathways. This intrinsic diversity makes them notoriously hard to control with single-pathway drugs, as some subsets of tumor cells inevitably evade any given treatment[111]. In such cases, a CKG-targeted therapy has maximal appeal: by hitting a common linchpin that all cancer subclones depend on (for example, a universal cell-cycle regulator or a metabolic enzyme critical under tumor conditions), one can cut across the genetic variability and impose a broad anti-tumor effect. Metastatic cancers are essentially system-level diseases, so targeting a system-critical target (the network hub) is a logical countermeasure. We may anticipate that cancers with high mutational burden or genomic instability will be especially reliant on keeper genes, as they need robust stress response and cell-cycle checkpoints to survive their chaotic genomes[114]. Indeed, tumors with many mutations like metastatic melanoma, lung cancer, or bladder cancer,could have pronounced dependencies on CKGs that manage DNA damage or proteotoxic stress[124]. Those tumor types might see significant benefit from adding a CKG inhibitor to standard care, either upfront or at progression, to achieve longer-lasting control[125].

On the other end of the spectrum, minimal residual disease (MRD) settings represent a prime opportunity for CKG-based interventions[119]. MRD exists in diseases like acute leukemia[126] or solid tumors after surgery. The cells in MRD are often under metabolic and oxidative stress and may be quiescent or slow-cycling, making them less susceptible to traditional cytotoxic drugs that target rapidly proliferating cells[127]. However, those cells still rely on keeper genes to survive in a dormant state[128]. Targeting a CKG in MRD could either flush these cells out of dormancy or directly induce their death/differentiation. The translational value here is enormous: eradicating MRD equates to curing the patient, as there would be no seeds left for relapse. For example, in chronic myeloid leukemia, maintenance with a tyrosine kinase inhibitor (targeting the driver *BCR-ABL*) has allowed some patients to achieve deep molecular remissions[129]; by analogy, in a solid tumor like triple-negative breast cancer, targeting a survival pathway like autophagy or a cell-cycle checkpoint active in residual micrometastases might prevent relapse after initial therapy[130]. Additionally, because MRD treatments are given when tumor burden is low, patients are more likely to tolerate them for extended periods as opposed to treating bulky disease, where rapid tumor kill is needed[131]. A well-chosen CKG inhibitor with a good safety profile could be administered for months or years as preventive therapy in high-risk patients, much like hormonal therapy is given for 5+ years in ER-positive breast cancer to suppress recurrence[132]. This could redefine adjuvant therapy by incorporating *network-targeted agents* that lock residual cancer cells into a non-proliferative state indefinitely[133].

Cancer types that might especially benefit from CKG-targeted strategies include those where current targeted therapies have shown only transient benefit due to resistance[134]. For instance, colorectal cancer with RAS mutations has limited targeted options. However, these tumors universally require robust cell-cycle progression and EGFR/MAPK signaling cross-talk – a network vulnerability that a CKG approach could exploit[135]. Pancreatic ductal adenocarcinoma, another lethal cancer with multiple resistance mechanisms, may rely on stress-response CKGs, such as regulators of reactive oxygen species or autophagy, to tolerate its harsh microenvironment; disabling such a CKG might render pancreatic cancer cells significantly less fit[136]. Even in cancers where targeted therapies have improved survival, like EGFR-mutant lung cancer or BRAF-mutant melanoma, CKG targeting could be deployed in later lines to prolong survival beyond what driver-centric therapies can achieve, by tackling the network adaptability that those tumors use to evade treatment.

**6. Challenges and Future Directions**

Although the CKG paradigm has opened new avenues for tumor intervention, it still faces multiple challenges that must be addressed in practice. Foremost, data quality imposes a critical bottleneck in constructing the gene-regulatory networks on which CKG inference depends. Existing repositories have catalogued numerous transcription-factor–target-gene regulatory edges. However, these documented interactions represent only a small fraction of the regulatory circuitry operating in vivo and thus fall far short of capturing the true complexity of cellular networks[137]. In particular, the absence of genome-wide, direct measurements of transcription-factor activity hampers the distinction between mere co-variation in gene expression and genuine causal regulation[138]. Such incompleteness and noise introduce biases into network models, thereby constraining the accuracy and reliability of CKG identification.

A second limitation arises from the prevalent "dynamic homogeneity" assumption embedded in many network-dynamical analyses, which presumes that distinct cell states share identical regulatory architectures and kinetic laws[139]. Conventional approaches, such as dynamic Bayesian networks, often treat network parameters as time-invariant, neglecting spatiotemporal fluctuations in regulatory processes and potentially yielding biased or erroneous conclusions[140]. In highly heterogeneous tumors, this static perspective may obscure critical network differences among cellular subpopulations, diminishing our understanding of intra-tumor heterogeneity.

Third, the limited cross-cancer transferability of current CKG methodologies restricts their translational scope. Strategies validated in a single tumor type frequently fail to generalize to other malignancies because each cancer harbors a unique regulatory topology and set of driver molecules[33]. Gene essentiality is markedly context-dependent, and ignoring this specificity risks serious misjudgment of gene indispensability[12]. Consequently, a CKG model optimized for one cancer may prove ineffective elsewhere, underscoring the need for methodological frameworks endowed with broader generalizability or applicability to tailor to specific cancer types.

To address these challenges, future CKG research must integrate multi-tier omics technologies with artificial intelligence (AI) methods, thereby enabling a systematic upgrade of CKG discovery and intervention strategies. Joint analysis of multi-omics data – transcriptomic, epigenomic, and proteomic, ideally at single-cell resolution – will substantially enhance the completeness and fidelity of gene-regulatory networks. The fusion of disparate omics layers can compensate for the limitations of any single data source: chromatin accessibility or transcription-factor binding derived from epigenomic assays can augment regulatory edges, while proteomic profiles can calibrate the activity states of key proteins, collectively yielding network models that better approximate biological reality. Concurrently, spatial-omics technologies, such as spatial transcriptomics and single-cell spatial atlases, provide a revolutionary means to dissect tumor spatial heterogeneity; analyses that disregard micro-environmental context often miss network disparities exhibited by the same cell type in different spatial niches[141]. Incorporating spatial information will allow precise localization of critical CKGs and their circuits within tissue architecture, elucidating micro-environmental influences on regulatory networks and guiding region-specific therapeutic strategies[142].

Moreover, the rapid evolution of AI technologies injects fresh impetus into CKG studies. Deep-learning models, particularly graph neural networks (GNNs), can seamlessly integrate graph structural representations with high-dimensional features, exhibiting superior pattern recognition and predictive performance on large-scale, multimodal biological data [143]. Unlike traditional black-box models, GNNs explicitly encode network topology, affording greater interpretability, and are intrinsically suited to fusing heterogeneous omics data[144,145]. Recent work has introduced attention-augmented, interpretable GNN frameworks to dissect cancer regulatory modules, highlighting key regulatory nodes and pathways through attention weights and thereby clarifying CKG mechanisms while bolstering biological credibility[146]. As multi-omics and spatial-omics datasets proliferate and AI models continue to advance, the CKG paradigm is poised to evolve from a static, unidimensional construct into a dynamic, holistic framework. Through the integrated application of these technologies, investigators can gain a more comprehensive understanding of the complexity inherent in cancer regulatory networks and, in turn, identify intervention targets and strategies that are both more universally applicable and have longer-lasting therapeutic efficacy.

## 7. Concluding Remarks

Cancer research is undergoing a decisive conceptual transition. The classical paradigm centered on cancer-driver mutations and oncogene addiction, while foundational, explains only part of tumor biology and has proved vulnerable to therapeutic resistance[147]. Insights from large-scale genomics, functional screens and systems modelling have redirected attention toward the mechanisms that maintain, rather than merely initiate, the malignant state[2]. The Cancer Keeper Gene (CKG) framework—defining genes that are indispensable for sustaining cancer irrespective of their mutational status—captures this shift and offers a unifying perspective on tumor dependency.

The CKG paradigm is theoretically innovative in several respects. First, by formally incorporating control theory in cancer biology, it shifts the focus from component-based to system-oriented thinking. Instead of identifying targets solely by mutation frequency or differential expression, it is now feasible to interrogate dynamic gene-regulatory networks to pinpoint control hubs whose perturbation can collapse malignant homeostasis. Second, the framework reconciles disparate dependency phenomena, such as oncogene addiction and non-oncogene addiction, by subsuming them under a single concept: any gene essential for the cancer attractor qualifies as a CKG. This extension breaks the long-standing binary division of tumor genes into "drivers" and "passengers," introducing a third category, namely "keepers," with a precise network-theoretic definition. Third, CKG research catalyzes interdisciplinary integration. Algorithms that exhaustively enumerate control hubs in polynomial time, graph-theoretical methods for constructing patient-specific regulatory networks, and machine-learning pipelines for inferring context-specific interactions exemplify the new toolkit arising from the convergence of graph theory, control science, bioinformatics and molecular oncology. These approaches not only advance cancer research but also illuminate other complex diseases, echoing the vision of "network medicine," in which disease is understood through the architecture of molecular interaction networks[14].

By foregrounding the systemic nature of malignancy, the CKG concept emphasizes that a tumor is a self-organized, rewired circuit whose stability arises from collective interactions. Understanding and disrupting cancer, therefore, requires identifying the fragile nodes that sustain this emergent state, rather than merely cataloging isolated mutations. Such a systems view helps bridge the gap between genomic alterations and phenotypic manifestation. For example, tumors harboring identical KRAS mutations often exhibit divergent therapeutic responses, partly because their internal network topology (and hence their set of CKGs) differs. Only by following the causal chain from mutation to network perturbation to phenotypic dependency can we rationally predict responses[103]. The CKG framework provides an analytical handle to quantify and model this chain, operationalizing systems-biology principles for translational use. More broadly, it expands the horizon of cancer biology, promoting the notion of network controllability and other system-level attributes as integral to precision medicine.

Ultimately, the CKG perspective points toward a therapeutic future guided not solely by mutational profiles but by a comprehensive understanding of the functional dependencies and network vulnerabilities that sustain each patient's tumor. Such insight will foster interventions designed to destabilize the malignant network as a whole, rather than targeting a single molecular pathway, thereby offering a plausible route to more durable and truly personalized cancer control.

**Author Contributions**

Xizhe Zhang and Weixiong Zhang conceived the research, drafted and revised the manuscript.

**Competing interests**

The authors declare that they have no competing interests.

**Funding**

This work was supported in part by the National Natural Science Foundation of China (Grant No. 62176129), National Natural Science Foundation of China-Jiangsu Joint Fund (Grant No. U24A20701), the Hong Kong RGC Strategic Target Grant (Grant No. STG1/M-501/23-N), the Hong Kong Health and Medical Research Fund (Grant No. 10211696), the Hong Kong RGC Theme-based Research Scheme (Grant No. TRS grant T24-508/22-N), NSFC/RGC Collaborative Research Scheme (Grant No. CRS_HKBU 2021/22), the Hong Kong RGC Collaborative Research Fund (Grant No. CRF C5005-23WF), the Hong Kong Global STEM Professorship Scheme, and the Hong Kong Jockey Club Charity Trust.